\documentstyle[12pt]{article}
\begin{document}
\newcommand{\ds}{\displaystyle}
\newcommand{\be}{\begin{equation}}
\newcommand{\ee}{\end{equation}}
\mbox{ }\hfill{\normalsize ITP-93-32E}\\
\mbox{ }\hfill{\normalsize hep-th/9306094}\\
\mbox{ }\hfill{\normalsize May 1993}\\
\begin{center}
{\Large \bf Exact Multiplicities in the Three-Anyon Spectrum}\\
\vspace{1cm}
{\large Stefan V.~Mashkevich\footnote{email: gezin@gluk.apc.org}}\\[.5cm]
{\large \it
N.N.Bogolyubov Institute for Theoretical Physics, \\ 252143 Kiev,
Ukraine}
\end{center}
\vspace{.5cm}
\begin{abstract}
Using the symmetry properties of the three-anyon spectrum, we obtain
exactly the multiplicities of states with given energy and angular
momentum. The results are shown to be in agreement with the proper
quantum mechanical and semiclassical considerations, and the
unexplained
points are indicated.
\end{abstract}
\newpage

     It is well known that the quantum mechanical spectrum of $N$
non-interacting bosons or  fermions  can  be  obtained  given  just
the
single-particle spectrum, but for anyons this does not hold  because
the $N$-anyon problem is essentially many-particle.  For  arbitrary
$N$
there are two classes of exact solutions \cite{Chou,Poly,Dunn,IJMP}
but
their relative
number decreases rapidly with $N$ increasing. The only case  in
which
one can proceed with the exact analysis more or less  far  ahead  is
that of $N=3.$ In the previous work \cite{PL} we have shown that it
is
possible to calculate exactly all the degeneracies in  the
three-anyon
spectrum using certain symmetry properties of the  latter.  Here  we
will carry out an analogous  calculation,  taking  into  account  in
addition the angular momentum. Our present consideration will  allow
us to shed at least some light on the problem of quantum  mechanical
description of anyonic spectra, which at  the  moment  is  far  from
being closed.

     As it was done earlier, we consider the problem of  three
non-interacting anyons in a harmonic potential, with the  particle
mass and the frequency set to unity. The Hamiltonian is
$\hat{H}=\sum_{j=1}^{3}\hat{H}_{j}$ with the one-particle Hamiltonian
$\hat{H}_{j}=\frac{1}{2}(-\Delta_{j}+{\bf r}_{j}^{2}).$
The single-particle state is uniquely determined by the two quantum
numbers -- energy $E$ which may equal $1,2,\dots,$ and angular
momentum
$L=-(E-1),-(E-3),\dots,E-3,E-1.$ Formally, the number of
single-particle
states with energy $E$ and momentum $L$ is given by
\be
g_{1}(E,L) = \left\{ \begin{array}{cl}
                r(E-L,2) & $ if $ |L|\le E-1 \\ \\
                0 & \mbox{ otherwise}
                \end{array}
             \right.
\label{single}
\ee
where $r(a,b)$ is the remainder of the division of $a$ by $b$. (Here
and
further $E$ is implicit to be a positive integer and $L$ an integer.)
In
what follows it will be also convenient to use, along with $E$ and
$L$,
the numbers ${\ell}^{+}=(E+L-1)/2$ and ${\ell}^{-}=(E-L-1)/2$.
Obviously,
any pair of non-negative integers $({\ell}^{+},{\ell}^{-})$
corresponds
to exactly one state.

     To calculate the multiplicities  $\tilde{g}_{3B}(E,L)$ and
$\tilde{g}_{3F}(E,L)$ in the relative motion spectrum of three
bosons  and  three  fermions,
respectively, we start from the multiplicities  $g_{3}(E,L)$
corresponding to Boltzmann statistics, then come to  $g_{3B}(E,L)$
and $g_{3F}(E,L)$ for the full spectrum, and afterwards  separate
the
center-of-mass motion.

     The calculation is simplified by using
\be
{\cal L}^{\pm}=\frac{\ds1}{\ds2}(E\pm L-3);
\label{lplus}
\ee
since ${\cal L}^{\pm}={\ell}^{\pm}_{1}+{\ell}^{\pm}_{2}+{\ell}^{\pm}
_{3}$ with the ${\ell}^{\pm}_{i}$'s as defined earlier, the
Boltzmann degeneracy is $g_{3}(E,L)=S_{3}({\cal L}^{+})S_{3}({\cal
L}^{-})$
where $S_{3}(N)$ is the number
of ordered triples of integers the sum of which is  $N$.
One has $S_{3}(N)=\frac{\ds1}{\ds2}(N+1)(N+2)$ and consequently
\be
%% FOLLOWING LINE CANNOT BE BROKEN BEFORE 70 CHAR
g_{3}(E,L)=\frac{1}{64}\left[(E^{2}-1)^{2}-2(E^{2}+1)L^{2}+L^{4}\right]. .
\ee
Now, in Boltzmann count one bosonic state is taken six times if  all
the particles occupy different single-particle states,  three  times
if two are in the same state but the third is in another one and one
time if all are in one and the same state. Therefore
\be
%% FOLLOWING LINE CANNOT BE BROKEN BEFORE 70 CHAR
g_{3B}(E,L)=\frac{1}{6}\left[g_{3}(E,L)+3d_{3}(E,L)+2t_{3}(E,L)\right].
\ee
where  $d_{3}(E,L)$ is the number of states with two  particles
in  the same state (the third one may be  or  not  be  in  that
state)  and  $t_{3}(E,L)$  is the number of states with all  three
particles  in  the same state. For fermions, correspondingly,
\be
%% FOLLOWING LINE CANNOT BE BROKEN BEFORE 70 CHAR
g_{3F}(E,L)=\frac{1}{6}\left[g_{3}(E,L)-3d_{3}(E,L)+2t_{3}(E,L)\right].
\ee
By virtue of the aforesaid,
\be
d_{3}(E,L)=Q_{3}({\cal L}^{+})Q_{3}({\cal L}^{-})
\ee
with $Q_{3}(N)$ the number of ordered pairs $(\ell_{1},\ell_{3})$
such that
$2\ell_{1}+\ell_{3}=N\;\;(\ell_{2}=\ell_{1}),$
\be
Q_{3}(N)=\left[\frac{N}{2}\right]+1\equiv\frac{1}{2}[N-r(N,2)]+1,
\ee
where $[n]$ stands for the entire part of $n$. Finally,
\be
t_{3}(E,L)=d(E,3)d(L,3),
\ee
where we have introduced the "multiplicity function"
\be
d(a,b)=
  \left\{ \begin{array}{cc}
     1 & $ if $ r(a,b)=0 \\ \\
     0 & \mbox{ otherwise}
     \end{array}
  \right.  .
\ee
Explicit expressions for $g_{3B}$ and $g_{3F}$ read
$$
%% FOLLOWING LINE CANNOT BE BROKEN BEFORE 70 CHAR
g_{3B}(E,L)=\frac{1}{384}\left[(E^{2}-L^{2})^{2}+10E^{2}-14L^{2}+24E+13\right]
$$
\be
+\frac{1}{16}[2{\cal M}^{+}{\cal M}^{-}+{\cal M}^{+}(L-E-1)-{\cal
M}^{-}(L+E+1)]+\frac{1}{3}d(E,3)d(L,3),
\label{g3b}
\ee
$$
%% FOLLOWING LINE CANNOT BE BROKEN BEFORE 70 CHAR
g_{3F}(E,L)=\frac{1}{384}\left[(E^{2}-L^{2})^{2}-14E^{2}+10L^{2}-24E-11\right]
$$
\be
-\frac{1}{16}[2{\cal M}^{+}{\cal M}^{-}+{\cal M}^{+}(L-E-1)-{\cal
M}^{-}(L+E+1)]+\frac{1}{3}d(E,3)d(L,3),
\label{g3f}
\ee
where ${\cal M}^{\pm}=r({\cal L}^{\pm},2)$ , ${\cal L}^{\pm}$ are
defined by
(\ref{lplus}), and $E$ and $L$ should be of different parity.

     In order to separate the center-of-mass motion, note that
\be
g_{3S}(E,L)=\sum_{e=1}^{E}\sum_{l=-e+1}^{e-1}\tilde{g}_{3S}(E-e,L-l)
\label{g3s}
\ee
for  $S = B,F$  ( $e$  and  $l$  are the center-of-mass energy and
angular momentum, respectively). Eq.(\ref{g3s}) implies
\be
\tilde{g}_{3S}(E,L)=g_{3S}(E+1,L)+g_{3S}(E-1,L)
-g_{3S}(E,L-1)-g_{3S}(E,L+1).
\label{gt3s}
\ee
Substituting (\ref{g3b}) and (\ref{g3f}), one gets
$$
\tilde{g}_{3(B/F)}(E,L)=\frac{1}{24}(E^{2}-L^{2})+\frac{1}{3}\left[
d(E+1,3)d(L,3)+d(E-1,3)d(L,3)\right.
$$
\be
%% FOLLOWING LINE CANNOT BE BROKEN BEFORE 70 CHAR
\left.-d(E,3)d(L-1,3)-d(E,3)d(L+1,3)\right]\pm\frac{1}{8}r(E-L,4)r(E+L,4),
\label{grel}
\ee
the upper/lower sign referring to  $B/F$ ; here,  $E$  and  $L$  have
 to
be of the same parity. In the table below, the values  of
$\tilde{g}_{3B}(E,L)$ and $\tilde{g}_{3F}(E,L)$  for all  $E\le 12$
are listed.

  Now, our main goal is to investigate how  the  spectrum
interpolates between the bosonic and fermionic  ones.  Anyonic wave
functions satisfy the interchange conditions  ${\cal P}_{jk}\Psi=
\exp(i\pi\delta)\Psi$ for each pair $j,k$ , where  ${\cal P}_{jk}$
is the operator of anticlockwise interchange
of particles  $j$  and  $k$  and  $\delta$  is the statistical
parameter; there is a continuous transition from bosons to
fermions as  $\delta$  goes  from 0 to 1 . Having turned an
$N$-anyon system by one complete  revolution
in an anticlockwise direction, one will have the wave function
multiplied by  $\exp[i\pi N(N-1)\delta]$ . Consequently, the
possible  values  of angular momentum for anyons are
$L = \frac{N(N-1)}{2}\delta+$integer. In  our case  $N = 3$
one gets  $L = 3\delta+$integer . Thus, a state with angular
momentum  $L$  at Bose statistics interpolates to  that  with
angular momentum  $L+3$  at Fermi statistics. As for energy,
it is known \cite{Spor,Murt,Illu} that possible values of
the difference  $E_{Fermi}-E_{Bose}$ are $+3,+1, -1,-3.$
In the first and the last cases the  $\delta$  dependence  of  $E$
is linear and the states may be found exactly (in  Ref.\cite{PL}  we
called them "good" states), in the other two ones ("bad" states)
this  dependence is non-linear and  not  found  exactly  (see
Ref.\cite{Amel} for an interesting hypothesis concerning the
mentioned dependence). To summarize, all bosonic states with the
quantum  numbers   $(E,L)$ fall into four classes according to
the  quantum  numbers  of  fermionic states to which they
interpolate;  those  can equal $(E+3,L+3),(E+1,L+3),(E-1,L+3),$ or
$(E-3,L+3).$ (It is implicit of course that in the subspace
of states with the same  $(E,L)$  one chooses the "correct" ones
in the same way as in perturbation theory with degenerate states;
it can always be done since  $E$  and  $L$  themselves are
good quantum numbers for anyons.)

     We will use  $\tilde{r}^{n}(E,L)$ to denote the number of
states which come from  $(E,L)$  at Bose statistics to  $(E+n,L+3)$
at Fermi  statistics. (Since it has been shown by numerical
calculations  \cite{Spor,Murt}  that  the energies of different
"bad" states interpolating  between  the  same bosonic and
fermionic ones are in general different  for  fractional
$\delta$ , one has for anyons, strictly speaking, "numbers" of
such  states rather than "multiplicities".) As  $n$ can take four
values, one needs four equations for each $(E,L)$ to determine those
numbers. The "law of conservation of states" at bosonic and
fermionic  points reads
\be
\tilde{r}^{+3}(E,L)+\tilde{r}^{+1}(E,L)+\tilde{r}^{-1}(E,L)+
\tilde{r}^{-3}(E,L)=\tilde{g}_{3B}(E,L),
\label{boson}
\ee
$$
\tilde{r}^{+3}(E-3,L-3)+\tilde{r}^{+1}(E-1,L-3)
$$
\be
+\tilde{r}^{-1}(E+1,L-3)+\tilde{r}^{-3}(E+3,L-3)
=\tilde{g}_{3F}(E,L),
\label{fermion}
\ee
respectively. Two more equations follow from the known symmetry
properties of the spectrum. First, in perturbation theory it is easy
to
establish \cite{SenN,SpoP} that at Fermi statistics, two states
with opposite angular momenta have opposite values of slopes, that
is,
derivatives  $(dE/d\delta)_{\delta=1}.$ Since the linear behavior
of  the  "good"  states  is exact, this implies the equality
\be
\tilde{r}^{+3}(E-3,L-3)=\tilde{r}^{-3}(E+3,-L-3)
\label{pert}
\ee
(see Fig.1). Second, there is the supersymmetry property. Namely,
it was pointed out by Sen \cite{SenP} that there exists an
operator  $\hat{Q}$  which annihilates some of the "good"
states but acting on  a  "bad"  state with statistical
parameter  $\delta$  always produces another  "bad"  state
with same energy  and  with  statistical  parameter  $1+\delta.$
It is straightforward to show that  $[\hat{L},\hat{Q}]=2\hat{Q}$
so that a state coming from  $(E,L)$  at  $\delta=0$ to $(E+1,L+3)$
at  $\delta=1$ turns  under  $\hat{Q}$ into that coming from
$(E,L+2)$  at $\delta=1$ to $(E+1,L+5)$ at $\delta=2.$
Now, parity transformation turns the latter into a state
coming from  $(E+1,-L-5)$  at  $\delta=0$  to $(E,-L-2)$
at $\delta = 1$  (Fig.2).  Therefore the last equation is
\be
\tilde{r}^{+1}(E,L)=\tilde{r}^{-1}(E+1,-L-5).
\label{super}
\ee
The four equations (\ref{boson})-(\ref{super}) and the expression
(\ref{grel}) for $\tilde{g}_{3B}$ and $\tilde{g}_{3F}$
would suffice to calculate  $\tilde{r}^{n}(E,L).$ It  seems,
however, hardly possible to obtain closed-form  expressions
for  them  in  a straightforward manner as it is not clear
in which form they  should be searched for. Instead, it is easy
to solve (\ref{boson})-(\ref{super}) numerically for low-lying
levels and it turns out to be possible to pick up the regularity
in the numbers.

    The results for  $\tilde{r}^{n}(E,L)$ for  $E\le 12$  are
summarized in Tab.2. The general tendency is clear and can be
said to coincide
with what one could more or less expect, but to move further it is
useful  to
involve the concept of towers \cite{Spor,SpoP,SenP}.  Again following
Ref.\cite{SenP},  we recall that there exists an operator
$\hat{K}_{-}$  such
that $[\hat{H},\hat{K}_{-}]=-2\hat{K}_{-}$ and
$[\hat{L},\hat{K}_{-}]=0$ ---
therefore it either annihilates a  common  eigenstate of $\hat{H}$
and
$\hat{L}$ or lowers  its  energy  by  two  units  without changing
its angular
momentum. This  means  in  turn  that  all the states fall into
"towers"
descending along which is realized by $\hat{K}_{-}$ until one reaches
a bottom
state for which $\hat{K}_{-}\Psi = 0.$ Therefore  any $(E,L)$ state
either
is  a  bottom  state  or  has  its  correspondent $(E-2,L)$ state
obtained from
it by action of $\hat{K}_{-}$ .  Denoting  by $\tilde{b}^{n}(E,L)$
the number
of bottom states with the  same characteristics as in
$\tilde{r}^{n}(E,L)$ ,
we have
\be
\tilde{b}^{n}(E,L)=\tilde{r}^{n}(E,L)-\tilde{r}^{n}(E-2,L),
\label{btor}
\ee and correspondingly
\be \tilde{r}^{n}(E,L)=\sum_{k=0}^{[E/2]}\tilde{b}^{n}(E-2k,L).
\label{rtob} \ee
Thus, instead of counting (and finding) all states it is sufficient
to count
(and find) only bottom states. The values of  $\tilde{b}^{+3}(E,L)$
and
$\tilde{b}^{+1}(E,L)$ are displayed in Tabs.3-4.

     Here at last the regularities are obvious. The  formal
expressions
read
$$
\tilde{b}^{+3}(E,L)=\left[\frac{E+1}{6}\right]-
\left[\frac{E+L}{4}\right]+\left[\frac{L}{2}\right]+1-d(E,6)
$$
\be
\mbox{for}\;\; \frac{E-2}{3}\le L\le E-2 ;
\label{bp3}
\ee
$$
\tilde{b}^{+1}(E,L)=\left[\frac{E^{*}}{6}\right]+
\left[\frac{L-L^{*}}{4}+\frac{1-r(E,2)}{2}\right]
+\frac{r(E^{*},6)}{4}
$$
$$
+\left[1-2r(E,2)\right]\frac{d(E^{*}-2,6)[r(L-L^{*},4)-1]}{2}
$$
\be
\mbox{for}\;\; \frac{-E-2}{3}\le L\le E-6 ,
\label{bp1}
\ee
where
\be
E^{*}=E-3r(E,2) , \;\;
L^{*}=2\left[\frac{E^{*}}{6}\right]-2+r(E,2) .
\label{eandl}
\ee
{}From these, one obtains the formulas for $\tilde{r}^{n}(E,L)$
immediately  by applying (\ref{rtob}):
\be
\tilde{r}^{+3}(E,L)=
       \left\{ \begin{array}{l}
         {\displaystyle
         \frac{L^{2}+6L+5}{12}+\frac{1-r(L,2)}{4}+\frac{d(L,3)}{3}}
\\  \\
         \hspace{5cm}$for $0\le L\le \frac{E-2}{3}, \\  \\
         {\displaystyle
         \frac{-E^{2}+6EL-5L^{2}+12E-12L}{48}+
         \frac{d(L,3)-d(E,3)}{3}} \\ \\
         +\left[{\textstyle 1-2r(E,2)}\right]
         {\displaystyle
         \frac{d(E-L-2,4)}{4}} \\  \\
         \hspace{5cm}$for $\frac{E-2}{3}\le L \le E-2;
       \end{array}\right.
\label{rp3}
\ee
\be
\tilde{r}^{+1}(E,L)=
         \left\{ \begin{array}{l}
         {\displaystyle
         \frac{E^{2}+6EL+9L^{2}+16E+48L+48}{48}+\frac{r(E-L,4)}{8}
         +\frac{d(E-1,3)}{3}}  \\  \\
         \hspace{6cm}$for $\frac{-E-2}{3}\le L <-1, \\ \\
         {\displaystyle
         \frac{E^{2}+6EL-15L^{2}+16E-72L-96}{48}+\frac{r(E-L,4)}{8}
         +\frac{d(E-1,3)}{3}}  \\  \\
         \hspace{6cm}$for $-1\le L \le\frac{E-8}{3}, \\  \\
         {\displaystyle
         \frac{E^{2}-2EL+L^{2}-4E+4L}{16}+\frac{r(E-L,4)}{8}} \\ \\
         \hspace{6cm}$for $\frac{E-8}{3}<L\le E -6.
       \end{array}\right.
\label{rp1}
\ee

Formulas (\ref{bp3}),(\ref{bp1}), (\ref{rp3}),(\ref{rp1}) are exact
and are
the main  result  of this paper. In these formulas,  $E$  and $L$
should  be  of  the  same parity, i.e.  $r(E-L,2) = 0$ ;
otherwise, as well as if  $L$ does  not fall in any of the ranges
specified in the formulas, the corresponding multiplicities
vanish. Formulas for  $\tilde{b}^{-1},\tilde{r}^{-1}$ and
$\tilde{b}^{-3},\tilde{r}^{-3}$ follow immediately from
the obtained ones upon applying (\ref{pert}) and (\ref{super}).

     Finally, summation over  $L$  yields the total numbers of
states with given energy and slope:
\be
\tilde{r}^{+3}(E)=\frac{1}{216}\left\{E^{3}+9E^{2}+
\left[42-27r(E,2)-24d(E,3)\right]E \\
+\left[f_{r(E^{*},6)}-81r(E,2)\right]\right\},
\ee
\be
\tilde{r}^{+1}(E)=\frac{1}{216}\left\{2E^{3}+3E^{2}+
\left[24d(E-1,3)-18\right]E \\
+\left[4r(E^{*},6)-27r(E,2)\right]\right\},
\ee
where  $f_{0}=0,f_{2}=88,f_{4}=56,$ and  $E^{*}$ has been defined
by (\ref{eandl}); these are exactly the expressions obtained in
Ref.\cite{PL} (where  slightly different notations were used).

     Let us now discuss the obtained results. For clarity, the
plots of  $\tilde{r}^{n}(E,L)$  for  $E = 100$  are displayed
in Fig.3. We  have  learned that for a given  $E$ , the states
of each of the four  classes  exist for not all values of  $L$
allowed for that  $E$ . Tab.5 shows the
lowest and the highest values of  $L$ for  which  the
corresponding states exist, as well as the values for which
the numbers of such states are maximal.

     Since all the states are solutions of the Schr\"{o}dinger
equation, all these values should in principle be deduced
directly  from  this equation. Producing them in this way is an
interesting and apparently difficult problem, and perhaps
at least some points of its  solution, if available, could
be used also to understand  the  structure of the  $N$-anyon
spectrum. What we can do at the moment is to explain
why for  $L=E-2$  and  $E-4$ , as the table shows,  there
exist  only the (+3) states. In the context of our previous
considerations, this follows immediately from (\ref{pert}) and
(\ref{super}) by noticing that $\tilde{r}^{+1}(E,E-k)
=\tilde{r}^{-1}(E+1,-E+k-5)$ does not vanish only for  $k\ge6$,
because  there are no $(E,L)$ states with  $L<-E+2$ , and
analogously for  $\tilde{r}^{-1}$ and $\tilde{r}^{-3}$ .
However, a purely quantum mechanical proof of this fact
can be given \cite{IJMP,UNP}, which we will describe here.

     Choosing as independent coordinates the one of the center
of mass $Z=(z_{1}+z_{2}+z_{3})/\sqrt{3}\;\;(z_{j}=x_{j}+iy_{j}$
is the complex coordinate of $j$-th particle) and two relative
ones  $z_{12}=(z_{1}-z_{2})/\sqrt{2}\;,\;z_{23}=(z_{2}-z_{3})/
\sqrt{2}$ and searching for the  wave  function
in the form  $\Psi=\Psi_{cm}\tilde{\chi}\exp\left[-\frac{1}{2}
(z_{1}z_{1}^{*}+z_{2}z_{2}^{*}+z_{3}z_{3}^{*})\right]$ , one
has for the relative Hamiltonian  $\tilde{H}=\tilde{H}_{(1)}
+\tilde{H}_{(2)}$  where $\tilde{H}_{(1)}=z_{12}\partial_{12}+
%% FOLLOWING LINE CANNOT BE BROKEN BEFORE 70 CHAR
z_{12}^{*}\partial_{12}^{*}+z_{23}\partial_{23}+z_{23}^{*}\partial_{23}^{*}
+2\;,\;\tilde{H}_{(2)}=-2(\partial_{12}\partial_{12}^{*}+
%% FOLLOWING LINE CANNOT BE BROKEN BEFORE 70 CHAR
\partial_{23}\partial_{23}^{*})\;,\;\partial_{jk}\equiv\partial/\partial z_{jk}
\;\;\;(\;\tilde{H}$ acts on $\tilde{\chi}$),
and for the relative angular momentum
$\tilde{L}=z_{12}\partial_{12}-
%% FOLLOWING LINE CANNOT BE BROKEN BEFORE 70 CHAR
z_{12}^{*}\partial_{12}^{*}+z_{23}\partial_{23}-z_{23}^{*}\partial_{23}^{*}.$
The function $\tilde{\chi}$ is to be searched  for
as a linear combination of functions of the form
\be
\left|l_{12}\;\bar{l}_{12}\;;\;l_{23}\;\bar{l}_{23}\;;\;
l_{31}\;\bar{l}_{31}\;\right\rangle_{+} \\
\equiv\left\{(z_{12})^{l_{12}}(z_{12}^{*})^{\bar{l}_{12}}
(z_{23})^{l_{23}}(z_{23}^{*})^{\bar{l}_{23}}
(z_{31})^{l_{31}}(z_{31}^{*})^{\bar{l}_{31}}\right\}_{+}
\label{func}
\ee
($z_{31}=-z_{12}-z_{23}$) where  $\{...\}_{+}$  means symmetrization
over  $z_{j}$'s   and the equalities  $l_{jk}-\bar{l}_{jk}=\delta
$+integer  should fulfil.  The  function (\ref{func})
then satisfies the anyonic interchange conditions
${\cal P}_{jk}\tilde{\chi}=\exp(i\pi\delta)\tilde{\chi},$
and the coefficients in the linear combination should
be chosen so that  $\tilde{\chi}$ be non-singular and satisfy
the equation $\tilde{H}\tilde{\chi}=E\tilde{\chi}$. Obviously,
$$
\tilde{L}\left|l_{12}\;\bar{l}_{12}\;;\;l_{23}\;\bar{l}_{23}\;;\;
l_{31}\;\bar{l}_{31}\;\right\rangle_{+}
$$
\be
=\left(l_{12}-\bar{l}_{12}+l_{23}-\bar{l}_{23}+
l_{31}-\bar{l}_{31}\right)
\left|l_{12}\;\bar{l}_{12}\;;\;l_{23}\;\bar{l}_{23}\;;\;
l_{31}\;\bar{l}_{31}\;\right\rangle_{+} ,
\ee
$$
\tilde{H}_{(1)}
\left|l_{12}\;\bar{l}_{12}\;;\;l_{23}\;\bar{l}_{23}\;;\;
l_{31}\;\bar{l}_{31}\;\right\rangle_{+}
$$
\be
=\left(l_{12}+\bar{l}_{12}+l_{23}+\bar{l}_{23}+
l_{31}+\bar{l}_{31}+2\right)
\left|l_{12}\;\bar{l}_{12}\;;\;l_{23}\;\bar{l}_{23}\;;\;
l_{31}\;\bar{l}_{31}\;\right\rangle_{+} ,
\ee
and $\tilde{H}_{(1)}\left|l_{12}\;\bar{l}_{12}\;;\;l_{23}\;
\bar{l}_{23}\;;\;l_{31}\;\bar{l}_{31}\;\right\rangle_{+} $ consists
of pieces of the form \[l_{jk}\bar{l}_{mn}\left|\dots l_{jk}-1\dots
\bar{l}_{mn}-1\dots\right\rangle_{+}.\]
The states which at  $\delta=0$
have the maximal angular momentum  $L = E-2$
are those with all  $\bar{l}_{jk}$ vanishing, i.e. of
the form  $\left|l\;0 ; m\;0 ; n\;0 \right\rangle_{+}$
(with integral  $l,m,n$);
the corresponding anyonic states  $\left|l+\delta\;0 ; m+\delta\; 0 ;
n+\delta\; 0 \right\rangle$ always are stationary because
$\tilde{H}_{(2)}$ annihilates  them,  and  their
energy is  $E=E_{Bose}+3\delta\;$(where  $E_{Bose}=l+m+n+2$).
Further, to construct a state which has $L=E-4$ at $\delta=0$
one should start from
$\left|l+\delta\; 1 ; m+\delta\;0 ; n+\delta\; 0 \right\rangle_{+}
\equiv\left|1\right\rangle$.
Acting on this with  $\tilde{H}_{(2)}$  yields
a sum of the terms of the form  $\left|p+\delta\; 0 ; q+\delta\; 0 ;
r+\delta\;0\right\rangle_{+}$, which are annihilated by
$\tilde{H}_{(2)}$ .
If all of them are non-singular,  then  a stationary
state can be built. (One has  $\tilde{H}_{(1)}\left|1\right\rangle=
E\left|1\right\rangle\;,\;
\tilde{H}_{(2)}\left|1\right\rangle=\left|2\right\rangle\;,\;
\tilde{H}_{(1)}\left|2\right\rangle=(E-2)\left|2\right\rangle\;,\;$
and $\tilde{H}_{(2)}\left|2\right\rangle=0$, consequently
$\tilde{H}\left[
\left|1\right\rangle+{\ds\frac{1}{2}}\left|2\right\rangle\right]=
%% FOLLOWING LINE CANNOT BE BROKEN BEFORE 70 CHAR
E\left[\left|1\right\rangle+{\ds\frac{1}{2}}\left|2\right\rangle\right]$.)
Now, the only potentially  singular  term in  $\left|2\right\rangle$
could be
$\left|-1+\delta\; 0 ; q+\delta\; 0 ; r+\delta\;0\right\rangle_{+}$
with  non-negative integral  $q$ and $r$  .  However,  such
function, which equals  $(z_{12}z_{23}z_{31})^{\delta}
\left| -1\; 0 ; q\; 0 ; r\;0 \right\rangle$  , can easily be shown to
be
non-singular: the singularities cancel out due to symmetrization
\cite{IJMP,UNP}. This completes the proof of the fact that
all bosonic states with $E=L-4$ , as well as those with $E=L-2$,
have their correspondent anyonic states with the slope +3.

     At present it is unknown how the other features of the
obtained results could be derived from  such  quantum
mechanical  considerations. However, the qualitative
picture can be justified using semiclassical arguments. In
the two-anyon problem,  there  is  only  one relative
coordinate  $z_{12}$ , the change of which is  governed  by  the
oscillator equation. The corresponding  classical
trajectories   are  ellipses,  and  the  anyonic  interchange
conditions  demand  that  the relative angular momentum be
$L=\delta+2l$ ,  where  the  integer $l$  determines the sign  of
$L$  at $\delta=0$ . Thus, as one starts   increasing $\delta$
from zero, one should expect linear increase of the energy of
those states in which the relative vector,  in  classical
terms, rotates anticlockwise $(l>0)$  or does not rotate
$(l=0)$ , and linear decrease for those in which it rotates
clockwise. This is indeed the case: The semiclassical
description of the two-anyon spectrum  turns out to yield the
exact values of levels \cite{Bhad,Illu}. For $N$ anyons,
there are $\frac{N(N-1)}{2}$ relative vectors, and the analogous
consideration would yield linear dependences with the  slopes
$E(1)-E(0)=\frac{N(N-1)}{2}-2s$ with $s$ the number of the
relative  vectors  rotating  clockwise. The set of the
slopes is correct \cite{Chin}, which encourages us  to use this
picture for a qualitative  analysis  (although  we  do  not
argue that it is good in other senses, for example that it
gives the correct multiplicities). So, for a state to  have
the  slope  +3  , neither  of  the  relative  vectors  should
rotate  clockwise.  Certainly it cannot be so if  $L<0$ .  For
$L=0$  a (+3)  state  can  be  only  realized  as  a  radial
excitation  of  the   bosonic   ground   state;   indeed,
$\tilde{r}^{+3}(E,0)=1$ (for even $E$ ). Further,  if  we
choose a state with a given $L$ "at random", then the more is
$L$ ,  the  more  vectors  "in average" rotate  anticlockwise.
Therefore the states with  $L$ close to maximal  $L=E-2$
have at most the slope +3 , and  with  $L$ decreasing,  states
with successively decreasing  slopes  emerge,  pass the
point  of their maximal number, and vanish, just as
one observes in Fig.3. Apparently the picture is analogous for
arbitrary $N$.

     To summarize, we have determined exactly the multiplicities  of
states with given energies and angular momenta in  the  spectrum  of
three non-interacting anyons in a harmonic well and shown  that  the
results are in conformity with certain quantum mechanical and
semiclassical considerations. Generally speaking, these  results
should follow directly from the Schr\"{o}dinger equation, but the
concrete procedure of deriving them this way is still to be found.

     I thank G.M.Zinovjev for his attention to my work and  constant
support, and Diptiman Sen for stimulating discussions. This work was
supported, in part, by a  Soros  Foundation  Grant  awarded  by  the
American Physical Society.

\newpage

\begin{center}
\begin{tabular}{|rr|c|c|c|rr|c|c|c|rr|c|c|c|rr|c|c|}
\cline{1-4}\cline{6-9}\cline{11-14}\cline{16-19}
$E$ & $L$ & $\tilde{g}_{3B}$ & $\tilde{g}_{3F}$ & \hspace{-.25cm} &
$E$ & $L$ & $\tilde{g}_{3B}$ & $\tilde{g}_{3F}$ & \hspace{-.25cm} &
$E$ & $L$ & $\tilde{g}_{3B}$ & $\tilde{g}_{3F}$ & \hspace{-.25cm} &
$E$ & $L$ & $\tilde{g}_{3B}$ & $\tilde{g}_{3F}$    \\
\cline{1-4}\cline{6-9}\cline{11-14}\cline{16-19}
2 & 0 & 1 & 0 & \hspace{-.25cm} & 7 & 1 & 2 & 2 &
\hspace{-.25cm} & 9 & -5 & 2 & 2 & \hspace{-.25cm} & 11 & -1 & 5 & 5
\\
\cline{1-4}
3 & 1 & 0 & 0 & \hspace{-.25cm} & 7 & -1 & 2 & 2 &
\hspace{-.25cm} & 9 & -7 & 1 & 1 & \hspace{-.25cm} & 11 & -3 & 5 & 5
\\
\cline{11-14}
3 & -1 & 0 & 0 & \hspace{-.25cm} & 7 & -3 & 2 & 2 &
\hspace{-.25cm} & 10 & 8 & 2 & 1 & \hspace{-.25cm} & 11 & -5 & 4 & 4
\\
\cline{1-4}
4 & 2 & 1 & 0 & \hspace{-.25cm} & 7 & -5 & 1 & 1 &
\hspace{-.25cm} & 10 & 6 & 3 & 3 & \hspace{-.25cm} & 11 & -7 & 3 & 3
\\
\cline{6-9}
4 & 0 & 1 & 1 & \hspace{-.25cm} & 8 & 6 & 2 & 1 &
\hspace{-.25cm} & 10 & 4 & 4 & 3 & \hspace{-.25cm} & 11 & -9 & 2 & 2
\\
\cline{16-19}
4 & -2 & 1 & 0 & \hspace{-.25cm} & 8 & 4 & 2 & 2 &
\hspace{-.25cm} & 10 & 2 & 4 & 4 & \hspace{-.25cm} & 12 & 10 & 2 & 1
\\
\cline{1-4}
5 & 3 & 1 & 1 & \hspace{-.25cm} & 8 & 2 & 3 & 2 &
\hspace{-.25cm} & 10 & 0 & 5 & 4 & \hspace{-.25cm} & 12 & 8 & 3 & 3
\\
5 & 1 & 1 & 1 & \hspace{-.25cm} & 8 & 0 & 3 & 3 &
\hspace{-.25cm} & 10 & -2 & 4 & 4 & \hspace{-.25cm} & 12 & 6 & 5 & 4
\\
5 & -1 & 1 & 1 & \hspace{-.25cm} & 8 & -2 & 3 & 2 &
\hspace{-.25cm} & 10 & -4 & 4 & 3 & \hspace{-.25cm} &
 12 & 4 & 5 & 5  \\
5 & -3 & 1 & 1 & \hspace{-.25cm} & 8 & -4 & 2 & 2 &
\hspace{-.25cm} & 10 & -6 & 3 & 3 & \hspace{-.25cm} & 12 & 2 & 6 & 5
\\
\cline{1-4}
6 & 4 & 1 & 0 & \hspace{-.25cm} & 8 & -6 & 2 & 1 &
\hspace{-.25cm} & 10 & -8 & 2 & 1 & \hspace{-.25cm} & 12 & 0 & 6 & 6
\\
\cline{6-9}\cline{11-14}
6 & 2 & 1 & 1 & \hspace{-.25cm} & 9 & 7 & 1 & 1 &
\hspace{-.25cm} & 11 & 9 & 2 & 2 & \hspace{-.25cm} & 12 & -2 & 6 & 5
\\
6 & 0 & 2 & 1 & \hspace{-.25cm} & 9 & 5 & 2 & 2 &
\hspace{-.25cm} & 11 & 7 & 3 & 3 & \hspace{-.25cm} & 12 & -4 & 5 & 5
\\
6 & -2 & 1 & 1 & \hspace{-.25cm} & 9 & 3 & 3 & 3 &
\hspace{-.25cm} & 11 & 5 & 4 & 4 & \hspace{-.25cm} & 12 & -6 & 5 & 4
\\
6 & -4 & 1 & 0 & \hspace{-.25cm} & 9 & 1 & 3 & 3 &
\hspace{-.25cm} & 11 & 3 & 5 & 5 & \hspace{-.25cm} & 12 & -8 & 3 & 3
\\
\cline{1-4}
7 & 5 & 1 & 1 & \hspace{-.25cm} & 9 & -1 & 3 & 3 &
\hspace{-.25cm} & 11 & 1 & 5 & 5 & \hspace{-.25cm} & 12 & -10 & 2 & 1
 \\
7 & 3 & 2 & 2 & \hspace{-.25cm} & 9 & -3 & 3 & 3 &
\hspace{-.25cm} & & & & & \hspace{-.25cm} & & & & \\
\cline{1-4}\cline{6-9}\cline{11-14}\cline{16-19}
\end{tabular}
\end{center}

Tab.1. The values of $\tilde{g}_{3B}(E,L)$ and $\tilde{g}_{3F}(E,L).$

\newpage

\begin{center}
\begin{tabular}{|rr|c|c|c|c|c|rr|c|c|c|c|c|rr|c|c|c|c|}
\cline{1-6}\cline{8-13}\cline{15-20}
 $E$ & $L$ & \multicolumn{4}{c|}{$n$} &  & $E$ & $L$ &
\multicolumn{4}{c|}{$n$}
 &  & $E$ & $L$ & \multicolumn{4}{c|}{$n$} \\
\cline{3-6}\cline{10-13}\cline{17-20}
   &  & 3 & 1 & -1 & -3 & & & & 3 & 1 & -1 & -3 &  &  &  & 3 & 1 & -1
& -3 \\
\cline{1-6}\cline{8-13}\cline{15-20}
 2  & 0 & 1 &  &  &  &  & 8 & 4 & 2 &   &   &   &  & 10 &-8 &  &  & 1
& 1  \\
\cline{1-6}\cline{15-20}
 3  & 1 &   &  &  &  &  & 8 & 2 & 2 & 1 &   &   &  & 11 & 9 & 2&  &
&    \\
 3  &-1 &   &  &  &  &  & 8 & 0 & 1 & 2 &   &   &  & 11 & 7 & 3&  &
&    \\
\cline{1-6}
 4  & 2 & 1 &  &  &  &  & 8 &-2 &   & 2 & 1 &   &  & 11 & 5 & 3& 1&
&    \\
 4  & 0 & 1 &  &  &  &  & 8 &-4 &   &   & 2 &   &  & 11 & 3 & 3& 2&
&    \\
 4  &-2 &   & 1&  &  &  & 8 &-6 &   &   & 1 & 1 &  & 11 & 1 & 1& 4&
&    \\
\cline{1-6}\cline{8-13}
 5  & 3 & 1 &  &  &  &  & 9 & 7 & 1 &   &   &   &  & 11 &-1 &  & 4& 1
&    \\
 5  & 1 & 1 &  &  &  &  & 9 & 5 & 2 &   &   &   &  & 11 &-3 &  & 2& 3
&    \\
 5  &-1 &   & 1&  &  &  & 9 & 3 & 2 & 1 &   &   &  & 11 &-5 &  &  & 4
&    \\
 5  &-3 &   &  & 1&  &  & 9 & 1 & 1 & 2 &   &   &  & 11 &-7 &  &  & 2
& 1  \\
\cline{1-6}
 6  & 4 & 1 &  &  &  &  & 9 &-1 &   & 3 &   &   &  & 11 &-9 &  &  & 1
& 1  \\
\cline{15-20}
 6  & 2 & 1 &  &  &  &  & 9 &-3 &   & 1 & 2 &   &  & 12 &10 & 2&  &
&    \\
 6  & 0 & 1 & 1&  &  &  & 9 &-5 &   &   & 2 &   &  & 12 & 8 & 3&  &
&    \\
 6  &-2 &   & 1&  &  &  & 9 &-7 &   &   & 1 &   &  & 12 & 6 & 4& 1&
&    \\
\cline{8-13}
 6  &-4 &   &  & 1&  &  &10 & 8 & 2 &   &   &   &  & 12 & 4 & 3& 2&
&    \\
\cline{1-6}
 7  & 5 & 1 &  &  &  &  &10 & 6 & 3 &   &   &   &  & 12 & 2 & 2& 4&
&    \\
 7  & 3 & 2 &  &  &  &  &10 & 4 & 3 & 1 &   &   &  & 12 & 0 & 1& 5&
&    \\
 7  & 1 & 1 & 1&  &  &  &10 & 2 & 2 & 2 &   &   &  & 12 &-2 &  & 4& 2
&    \\
 7  &-1 &   & 2&  &  &  &10 & 0 & 1 & 4 &   &   &  & 12 &-4 &  & 1& 4
&    \\
 7  &-3 &   & 1& 1&  &  &10 &-2 &   & 3 & 1 &   &  & 12 &-6 &  &  & 4
& 1  \\
 7  &-5 &   &  & 1&  &  &10 &-4 &   & 1 & 3 &   &  & 12 &-8 &  &  & 2
& 1  \\
\cline{1-6}
 8  & 6 & 2 &  &  &  &  &10 &-6 &   &   & 2 & 1 &  & 12 &10 &  &  & 1
& 1  \\
\cline{1-6}\cline{8-13}\cline{15-20}
\end{tabular}
\end{center}

Tab.2. The values of  $\tilde{r}^{n}(E,L).$ Zeros are not
written down for clarity.

\newpage

\begin{center}
\begin{tabular}{||c|r|cccccccccc||}
\cline{1-12}
\multicolumn{2}{||c|}{} &
\multicolumn{10}{c||}{$E$}  \\
\cline{3-12}
\multicolumn{2}{||c|}{} & 2 & 4 & 6 & 8 & 10 & 12 & 14 & 16 & 18 & 20
 \\
\cline{1-12}
  & 0 & 1 & 0 & 0 & 0 & 0 & 0 & 0 & 0 & 0 & 0 \\
  & 2 &  & 1 & 0 & 1 & 0 & 0 & 0 & 0 & 0 & 0 \\
  & 4 &  &  & 1 & 1 & 1 & 0 & 1 & 0 & 0 & 0  \\
  & 6 &  &  &  & 2 & 1 & 1 & 1 & 1 & 0 & 1  \\
$L$ & 8 &  &  &  &  & 2 & 1 & 2 & 1 & 1 & 1  \\
  & 10 &  &  &  &  &  & 2 & 2 & 2 & 1 & 2 \\
  & 12 &  &  &  &  &  &  & 3 & 2 & 2 & 2 \\
  & 14 &  &  &  &  &  &  &  & 3 & 2 & 3  \\
  & 16 &  &  &  &  &  &  &  &  & 3 & 3  \\
  & 18 &  &  &  &  &  &  &  &  &  & 4 \\
\cline{1-12}
\end{tabular}

\begin{tabular}{||c|r|cccccccccc||}
\cline{1-12}
\multicolumn{2}{||c|}{} &
\multicolumn{10}{c||}{$E$}  \\
\cline{3-12}
\multicolumn{2}{||c|}{} & 3 & 5 & 7 & 9 & 11 & 13 & 15 & 17 & 19 & 21
 \\
\cline{1-12}
  & 1 & 0 & 1 & 0 & 0 & 0 & 0 & 0 & 0 & 0 & 0 \\
  & 3 &  & 1 & 1 & 0 & 1 & 0 & 0 & 0 & 0 & 0 \\
  & 5 &  &  & 1 & 1 & 1 & 1 & 0 & 1 & 0 & 0  \\
  & 7 &  &  &  & 1 & 2 & 1 & 1 & 1 & 1 & 0  \\
$L$ & 9 &  &  &  &  & 2 & 2 & 1 & 2 & 1 & 1  \\
  & 11 &  &  &  &  &  & 2 & 2 & 2 & 2 & 1 \\
  & 13 &  &  &  &  &  &  & 2 & 3 & 2 & 2 \\
  & 15 &  &  &  &  &  &  &  & 3 & 3 & 2  \\
  & 17 &  &  &  &  &  &  &  &  & 3 & 3  \\
  & 19 &  &  &  &  &  &  &  &  &  & 3 \\
\cline{1-12}
\end{tabular}

Tab.3. The values of  $\tilde{b}^{+3}(E,L)$ .
\end{center}

\newpage

\begin{center}
\begin{tabular}{||c|r|cccccccccc||}
\cline{1-12}
\multicolumn{2}{||c|}{} &
\multicolumn{10}{c||}{$E$}  \\
\cline{3-12}
\multicolumn{2}{||c|}{} & 2 & 4 & 6 & 8 & 10 & 12 & 14 & 16 & 18 & 20
 \\
\cline{1-12}
  & -6& 0 & 0 & 0 & 0 & 0 & 0 & 0 & 1 & 0 & 1 \\
  & -4& 0 & 0 & 0 & 0 & 1 & 0 & 1 & 1 & 1 & 1 \\
  & -2&   & 1& 0 & 1 & 1 & 1 & 1 & 2 & 1 & 2  \\
  & 0 &   &  & 1& 1 & 2 & 1 & 2 & 2 & 2 & 2  \\
  & 2 &   &  &  & 1 & 1 & 2 & 2 & 3 & 2 & 2  \\
$L$ & 4  &  &  &  &  & 1 & 1 & 2 & 2 & 3 & 3  \\
  &  6 &   &  &  &  &  & 1 & 1 & 2 & 2 & 3 \\
  &  8 &   &  &  &  &  &  & 1 & 1 & 2 & 2 \\
  & 10 &   &  &  &  &  &  &  & 1 & 1 & 2  \\
  & 12 &   &  &  &  &  &  &  &  & 1 & 1  \\
  & 14 &   &  &  &  &  &  &  &  &  & 1 \\
  & 16 &   &  &  &  &  &  &  &  &  &   \\
\cline{1-12}
\end{tabular}

\begin{tabular}{||c|r|cccccccccc||}
\cline{1-12}
\multicolumn{2}{||c|}{} &
\multicolumn{10}{c||}{$E$}  \\
\cline{3-12}
\multicolumn{2}{||c|}{} & 3 & 5 & 7 & 9 & 11 & 13 & 15 & 17 & 19 & 21
 \\
\cline{1-12}
  & -7& 0 & 0 & 0 & 0 & 0 & 0 & 0 & 0 & 1 & 0 \\
  & -5& 0& 0 & 0 & 0 & 0 & 1 & 0 & 1 & 1 & 1 \\
  & -3& 0& 0& 1 & 0 & 1 & 1 & 1 & 1 & 2 & 1  \\
  & -1&  & 1& 1& 1 & 1 & 2 & 1 & 2 & 2 & 2  \\
  &  1&  &  & 1& 1 & 2 & 2 & 2 & 2 & 3 & 2  \\
$L$ & 3 &  &  &  & 1& 1 & 2 & 2 & 3 & 3 & 3  \\
  &  5 &  &  &  &  & 1& 1 & 2 & 2 & 3 & 3 \\
  &  7 &  &  &  &  &  & 1& 1 & 2 & 2 & 3 \\
  &  9 &  &  &  &  &  &  & 1& 1 & 2 & 2  \\
  & 11 &  &  &  &  &  &  &  & 1& 1 & 2  \\
  & 13 &  &  &  &  &  &  &  &  & 1& 1 \\
  & 15 &  &  &  &  &  &  &  &  &  & 1 \\
\cline{1-12}
\end{tabular}

Tab.4. The values of  $\tilde{b}^{+1}(E,L)$ .
\end{center}

\newpage

\begin{center}
\begin{tabular}{|c|c|c|c|}
\hline
    & Low & Max. & High \\ \hline
$ +3$ & 0 & ${\ds\frac{3E-6}{5}}$ & $E-2$ \\ \hline
$ +1$ & ${\ds\frac{-E-2}{3}}$ & ${\ds\frac{E-12}{5}}$ & $E-6$ \\
\hline
$ -1$ & $-E+2$ & ${\ds\frac{-E-12}{5}}$ & ${\ds\frac{E-14}{3}}$ \\
\hline
$ -3$ & $-E+2$ & ${\ds\frac{-3E-6}{5}}$ & $-6$ \\ \hline
\end{tabular}
\end{center}

Tab.5. The lowest and highest values of $L$ for which the
states belonging to $\tilde{r}^{n}(E,L)$ exist, and the
value for which the number of such states is maximal.

\newpage

{\large Figure Captions}

     Fig.1. Two "good" states with opposite angular momenta
            in the fermion limit.

     Fig.2. Two "bad" states related by (supersymmetry+parity)
            transformation.

     Fig.3. The plots of  $\tilde{r}^{n}(E,L)$  for  $E = 100$ .


\newpage

\begin{thebibliography}{99}
\bibitem{Chou}C.Chou,  {\em Phys.Rev.} D {\bf 44}, 2533 (1991);
Erratum:
Phys.Rev. D 45, 1433 (1992).
\bibitem{Poly}A.P.Polychronakos,  {\em Phys.Lett.} {\bf B264}, 362
(1991).
\bibitem{Dunn} G.Dunne, A.Lerda, S.Sciuto, C.A.Trugenberger,  {\em
Nucl.Phys.}
{\bf B370}, 601 (1992).
\bibitem{IJMP}S.V.Mashkevich,  {\em Int.J.Mod.Phys.} {\bf A7}, 7931
(1992).
\bibitem{PL}S.V.Mashkevich,  {\em Phys.Lett.} {\bf B295}, 233 (1992).
\bibitem{Spor}M.Sporre, J.J.M.Verbaarschot, I.Zahed,  {\em
Phys.Rev.Lett.}
{\bf 67}, 1813 (1991).
\bibitem{Murt}M.V.N.Murthy, J.Law, M.Brack, R.K.Bhaduri,
{\em Phys.Rev.Lett.}  67, 1817 (1991);  {\em Phys.Rev.} B {\bf 45},
4289 (1992).
\bibitem{Illu}F.Illuminati, F.Ravndal, J.Aa.Ruud,  {\em Phys.Lett.}
{\bf A161}, 323 (1992).
\bibitem{Amel}C.Chou, L.Hua, G.Amelino-Camelia,  {\em Phys.Lett.}
{\bf B286}, 329 (1992);
see also  G.Amelino-Camelia,  {\em Phys.Lett.} {\bf B299}, 83 (1993).
\bibitem{SenN}D.Sen,  {\em Nucl.Phys.} {\bf B360}, 397 (1991).
\bibitem{SpoP}M.Sporre, J.J.M.Verbaarschot, I.Zahed,  preprint
SUNY-NTG-91/40
(1991).
\bibitem{SenP}D.Sen,  {\em Phys.Rev.Lett.} {\bf 68}, 2977 (1992).
\bibitem{UNP}S.V.Mashkevich  (unpublished).
\bibitem{Bhad}R.K.Bhaduri, R.S.Bhalerao, A.Khare, J.Law,
M.V.N.Murthy,
{\em Phys.Rev.Lett.} {\bf 66}, 523 (1991).
\bibitem{Chin}S.A.Chin, C.-R.Hu,  {\em Phys.Rev.Lett.} {\bf 69}, 229
(1992).
\end{thebibliography}
\end{document}